\documentclass[referee]{raa}            

\usepackage{graphicx,times}             
\usepackage{natbib}                     
\usepackage{amssymb,amsmath}              
\bibpunct{(}{)}{;}{a}{}{,}                

\usepackage[pagebackref=true]{hyperref}  
\usepackage{multirow}

\begin{document}

  \title{Machine Learning-Based Identification of Contaminated Images in Light Curves Data Preprocessing
}

   \volnopage{Vol.0 (20xx) No.0, 000--000}      
   \setcounter{page}{1}          

   \author{Hui Li  
      \inst{1,2}
   \and Rong-Wang Li
      \inst{1,3}
   \and Peng Shu
      \inst{1}
   \and Yu-Qiang Li
      \inst{1,3}
   }
   \institute{Yunnan Observatories, Chinese Academy of Sciences, Kunming 650216, China; {\it lihui@ynao.ac.cn}\\
   \and
   University of Chinese Academy of Sciences, Beijing 100049, China
   \and
   Key Laboratory of Space Object and Debris Observation,Chinese Academy of Sciences,Nanjing 210023, China
\vs\no
   }

\abstract{ 
   Attitude is one of the crucial parameters for space objects and plays a vital role in collision prediction and debris removal.
    Analyzing light curves to determine attitude is the most commonly used method.
    In photometric observations, outliers may exist in the obtained light curves due to various reasons.
    Therefore, preprocessing is required to remove these outliers to obtain high quality light curves.
    Through statistical analysis, the reasons leading to outliers can be categorized into two main types: first, the brightness of the object significantly 
     increases due to the passage of a star nearby, referred to as ``stellar contamination," and second,
      the brightness markedly decreases due to cloudy cover, 
     referred to as ``cloudy contamination." Traditional approach of manually inspecting images for contamination is time-consuming and labor-intensive. 
    However, We propose the utilization of machine learning methods as a substitute. Convolutional Neural Networks (CNN) and Support Vector Machines 
     (SVM) are employed to identify cases of stellar contamination and cloudy contamination, achieving F1 scores of 1.00 and 0.98 on test set, 
     respectively. We also explored other machine learning methods such as Residual Network-18 (ResNet-18) and Light Gradient
     Boosting Machine (lightGBM), then conducted comparative analyses of the results.
}

   \authorrunning{H. Li, R.-W. Li, P. Shu, \& Y.-Q. Li }            
   \titlerunning{Machine Learning-Based Identification}  
 
   \maketitle

\section{Introduction}        
\label{sect:intro}
\noindent Light curves refers to the curves depicting changes in luminosity. 
The photometry of space objects is influenced by various factors, including the 
geometry involving the Sun, the space object, and the observer, as well as the 
object's shape, orientation, and surface reflectance characteristics.
Light curves are essential for studying the rotation state and characteristics 
of space objects. However, before obtaining light curves, it is necessary to preprocess
 the source data, which includes the removal of outliers and data contaminated by either star or cloud. 
 This preprocessing step ensures the acquisition of high-quality observational data.

\noindent The usual preprocessing method often requires manual judgment.
But when dealing with large volume of data, this judgment is time-consuming 
and labor-intensive. Using machine learning for pattern recognition can significantly 
improve efficiency and save a substantial amount of time and effort. 
Machine learning has widespread applications in Astronomy, including but not 
limited to predicting atmospheric seeing in optical observations (\citealt{Ni+etal+2022}), 
identifying AGN and pulsar candidates (\citealt{Zhu+etal+2021}), detecting outliers
 in astronomical images (\citealt{Han+etal+2022}), and classifing Gaia data (\citealt{Bai+etal+2018}).

\noindent \cite{Hinton+Salakhutdinov+2006} published a paper with two main points: (1) Artificial
 neural networks with multiple hidden layers exhibit exceptional feature learning 
 capabilities. (2) The effective overcoming of training difficulties in deep neural 
 networks can be achieved through ``layerwise pre-training," which introduced the field 
 of deep learning (\citealt{Zhou+2017}). In fact, there were even highly efficient deep 
 learning models proposed before 2006, such as CNN. In the 1980s 
 and 1990s, some researchers published studies on CNN in the field of pattern recognition, 
 showing excellent performance in handwritten digit recognition (\citealt{Lawrence+1997}, \citealt{Neubauer+1998}). However, at that 
 time, CNN still performed poorly with large-scale data. It wasn't until 2012 that 
 \cite{Krizhevsky+2012} used an extended deep CNN to achieve the best classification 
 results in the ImageNet Large Scale Visual Recognition Challenge (LSVRC), which brought
  CNN into the limelight and gained increasing attention from researchers.

\noindent The cloud detection method has made significant breakthroughs in remote sensing and Meteorology, 
with various theories and approaches proposed. Existing algorithms primarily focus on utilizing 
cloud spectral information, frequency data, spatial textures, and combine methods such as 
thresholding, clustering, artificial neural networks, and support vector machines for cloud detection (\citealt{Zhang+2018}).
 In the field of Astronomy, \cite{Mommert+2020} used ResNet-18 and lightGBM to determine the 
 presence of clouds, and \cite{Wang+2019} employed support vector machines 
 to assess cloud's presence. Both lightGBM and SVM demonstrated high discrimination accuracy. Accuracy denotes the proportion
  of correctly predicted samples to the total number of samples in classification tasks in this paper.

\noindent The purpose of this study is to use machine learning models to identify stellar contamination images and cloudy contamination images. 
A CNN is employed for the binary classification
 of stellar contamination images and norm images in this study.  
 Three methods including the lightGBM model, SVM, and ResNet-18
  are utilized for the classification of ``cloudy contamination". These classification techniques have been adapted to suit the requirements of our study.

\section{Data}
\label{sect:data}
\noindent 
Data of space objects, including satellites and space debris, observed in Yunnan 
Astronomical Observatory(YNAO)'s 1.2-meter telescope.
The telescope's field of view is $36'\times36'$, and the CCD model is 
[Andor] DU888\_BV(10687) with a size of $1024\times1024$ pixels.

\noindent Approximately 1,000,000 FIT (Flexible Image Transport System) images, taken of space objects, were obtained in Yunnan Astronomical Observatory during 2022. Manual labeling was performed on partly observational 
data from 2022, resulting in around 30,000 images labeled as ``normal," 1756 images labeled as ``cloud," 
and 582 images labeled as ``star." The label ``star" indicates stellar contamination, while the label 
``cloud" indicates cloudy contamination. These labeled images are stored in an SQL (Structured Query Language) file, and they can be directly retrieved by using the corresponding label.

\begin{figure}[h]
   \begin{minipage}[t]{0.33\linewidth}
   \centering
    \includegraphics[width=60mm]{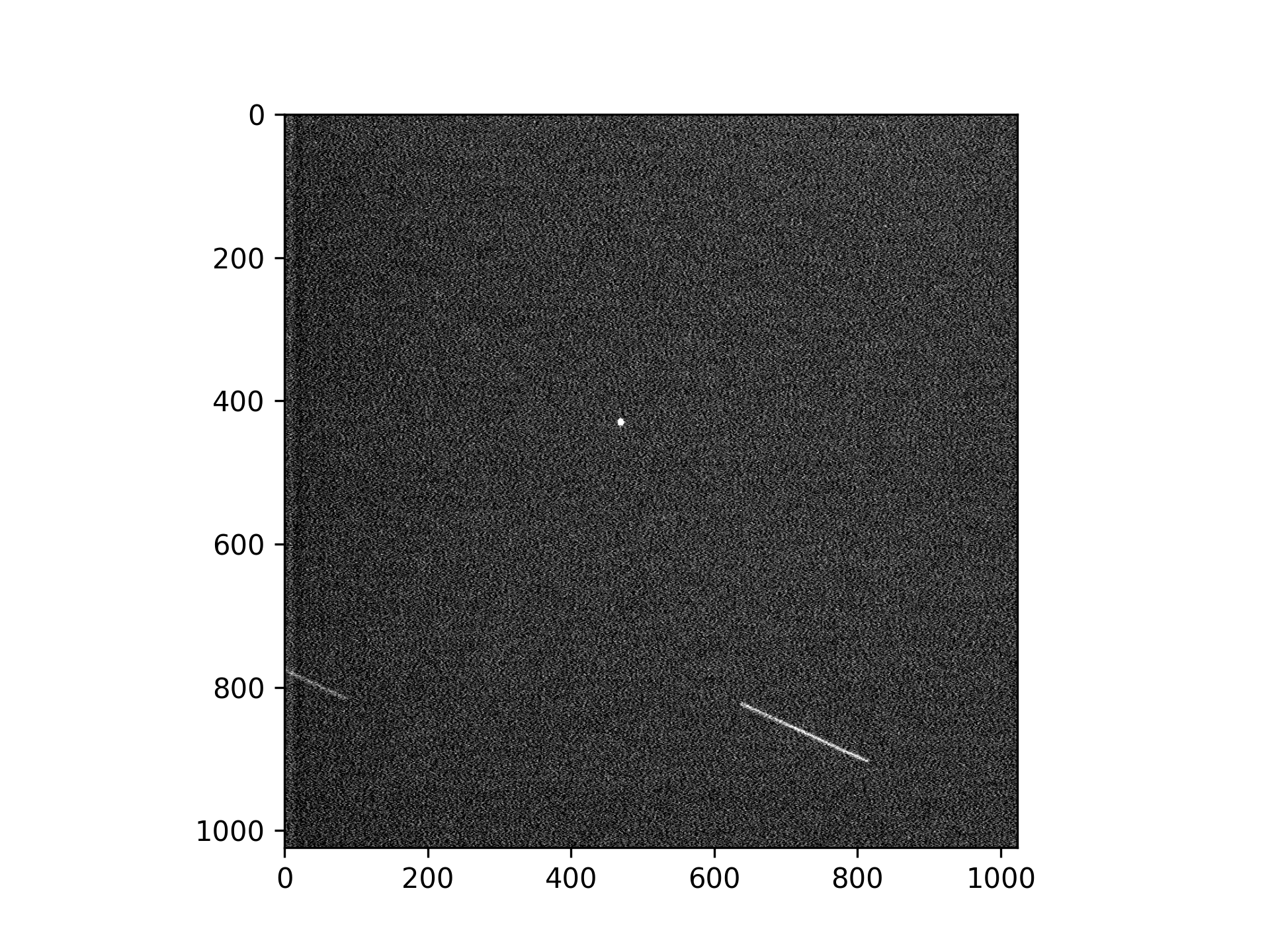}
      \caption{\label{Fig01}{\small normal} }
   \end{minipage}%
   \begin{minipage}[t]{0.33\textwidth}
   \centering
    \includegraphics[width=60mm]{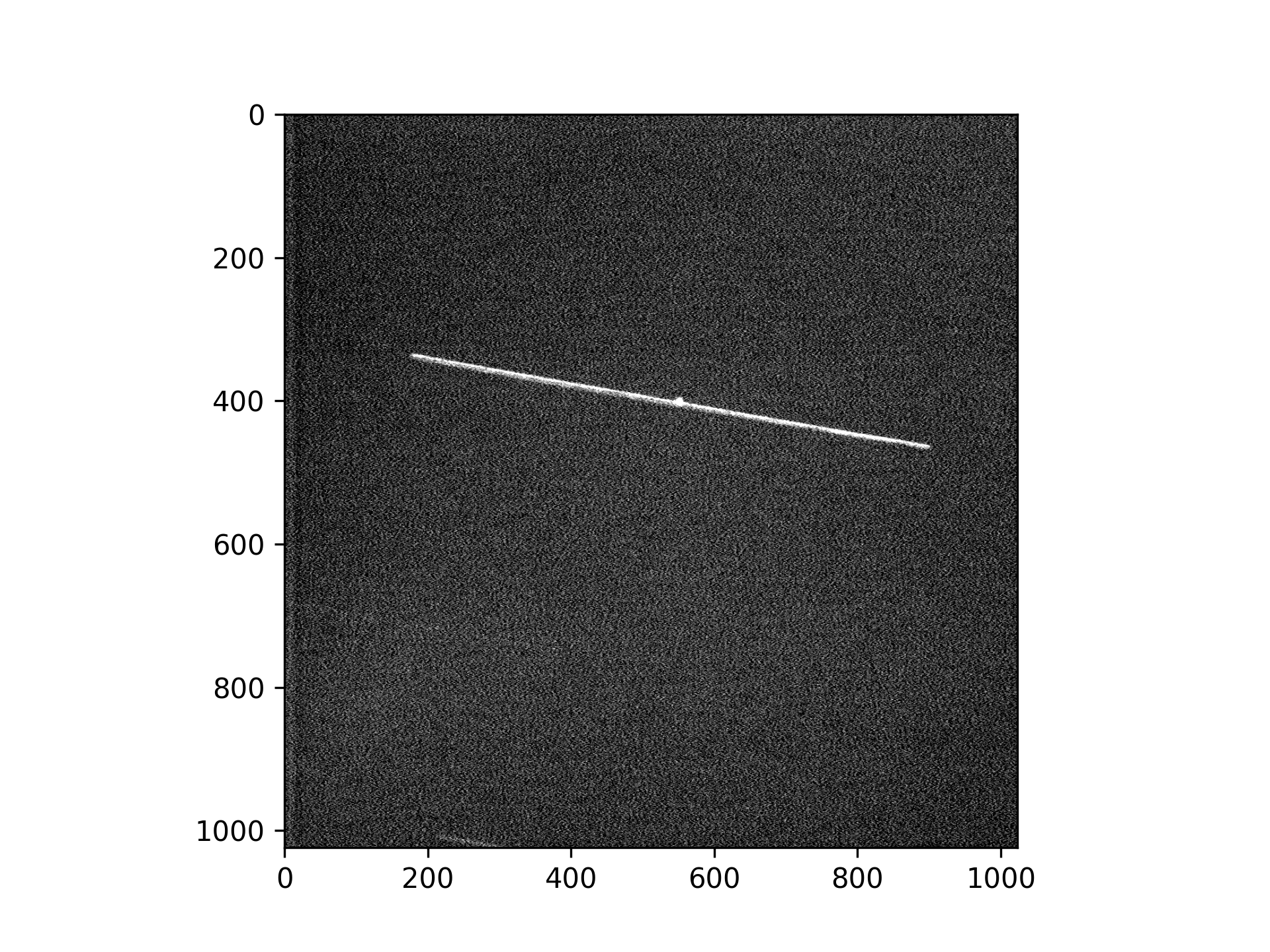}
      \caption{\label{Fig02}{\small star}}
   \end{minipage}%
   \begin{minipage}[t]{0.33\textwidth}
      \centering
       \includegraphics[width=60mm]{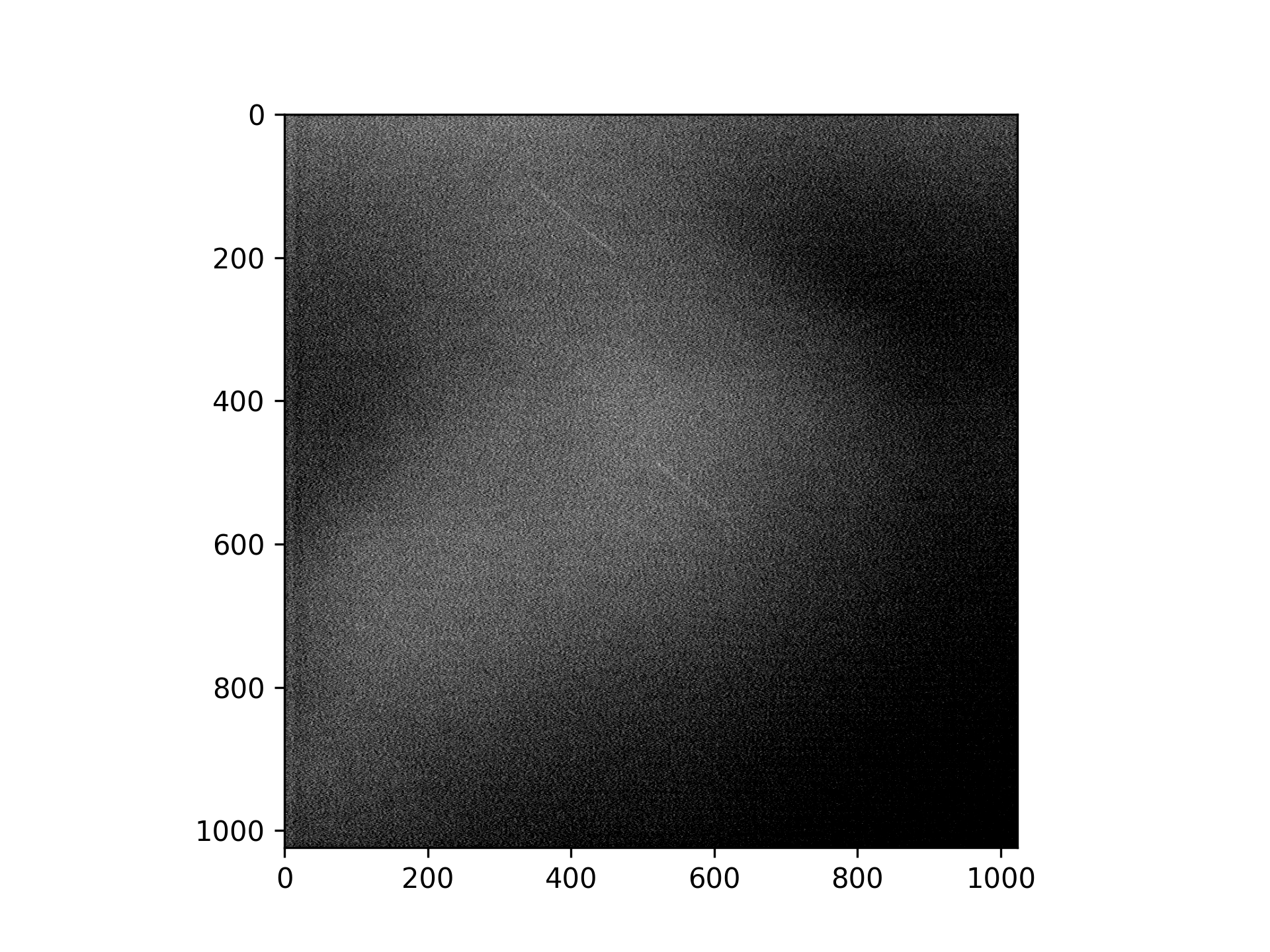}
         \caption{\label{Fig03}{\small cloud}}
      \end{minipage}%
 \end{figure}

\noindent 
By removing images contaminated by star and cloud, the data quality can be improved for further research and analysis.
The following 3 images, Fig.~\ref{Fig01}, Fig.~\ref{Fig02}, Fig.~\ref{Fig03}, represent 3 classes mentioned above.

\begin{figure}[h]
   \centering
   \includegraphics[width=60mm]{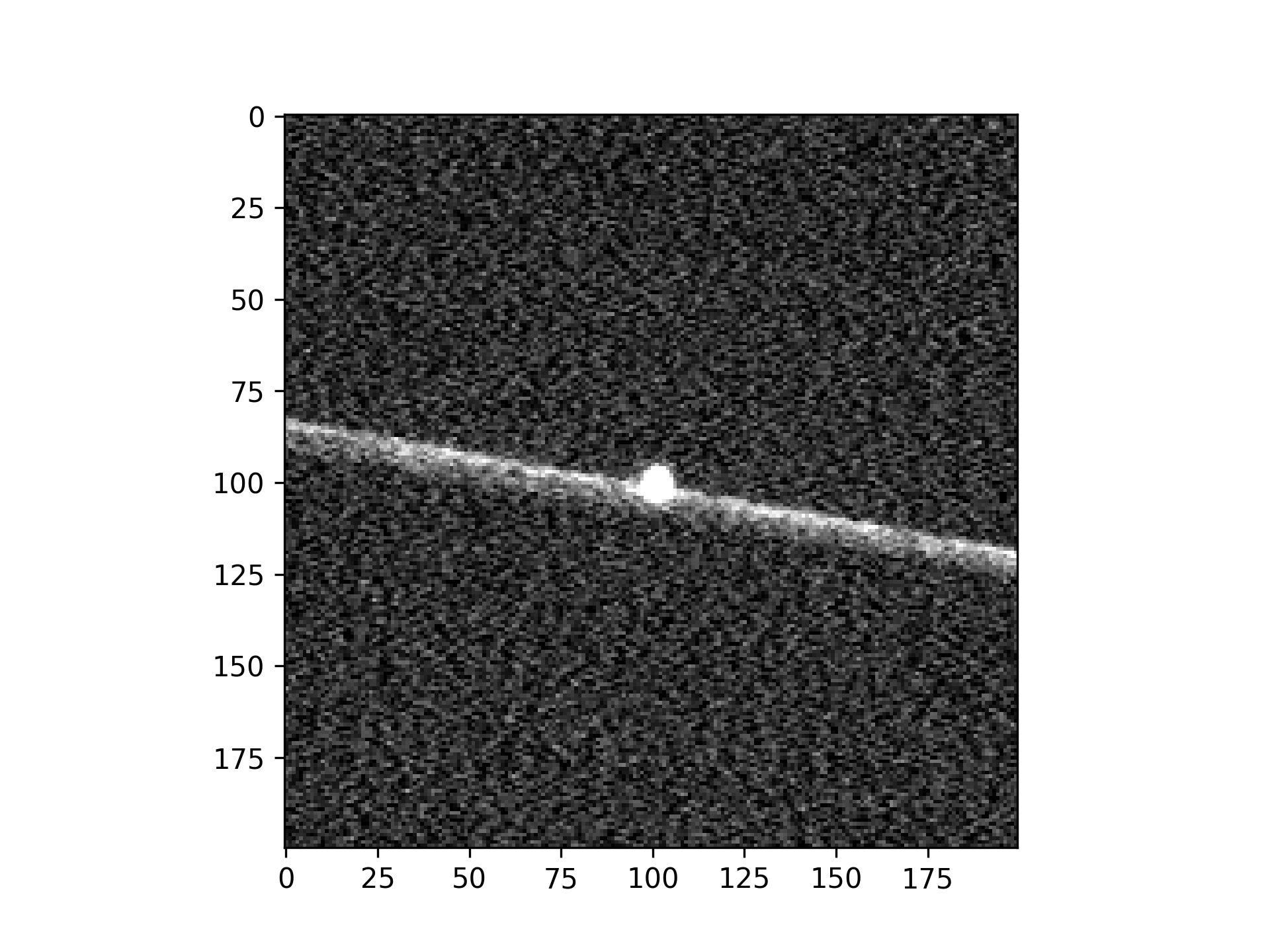}
   \caption{stamp image captured from Fig.2}
\label{Fig04}
\end{figure}

\noindent A CNN
 was employed as a binary classification approach to classify stellar contamination images. This model achieved a classification
  success rate of 92.21\% on the test set. Analysis suggests that image features 
  are related to space object and elongated star, which makes it more reasonable to perform 
  classification in partial region including space object. 
  Therefore, an attempt was made to process the data in the ``star" class to obtain stamp
   images with size of $200\times200$ pixels. The moving target is located at the center of stamp image (as shown in the Fig.~\ref{Fig04} captured from Fig.~\ref{Fig02}). A CNN model
    trained on stamp images achieved 100\% accuracy on the test set.

\noindent  It was found that images' overall standard deviation increased when images contained very bright
objects or stars, during the labeling process. This could lead to misclassifying normal images as having cloudy
  contamination. To ensure the quality of the dataset and prevent such misclassifications 
  during labeling, images were directly captured by the telescope in moving-target-tracking
  mode during nighttime. Cloud-contaminated images at two nights (October 4th and 12nd, 2023) were captured.

\noindent Finally, a machine learning dataset was established, which included ``normal" labeled data, 
 ``cloudy contamination" data, and the ``stellar 
contamination" data  (augmented by rotating and flipping). The dataset sizes and how many images used in each model are as shown in 
Tab.~\ref{Tab1}.
It was unable to extract the required features from some images during data preparation, 
resulting in differences in the dataset sizes used by different models.
\begin{table}[h]
   \centering
   \caption[]{Subtotals of all classes in the dataset and images used in each model}\label{Tab1}
   \begin{tabular}{ccccccc}
   \hline \noalign{\smallskip}
   \multirow{2}*{Classification} & \multirow{2}*{Count} & \multicolumn{5}{c}{Used in}  \\ \noalign{\smallskip} \cline{3-7} \noalign{\smallskip}
   &                    & CNN & CNN(stamp images)& ResNet & lightGBM & SVM  \\ \noalign{\smallskip} \hline 
   normal               & 31558 &4600 &2448 &15000 &10334 &15000   \\
   cloudy contamination (Oct 4th)   & 3199  &0    &0    &3199  &2527 &3199 \\
   cloudy contamination (Oct 12nd)  & 302   &0    &0    &302   &231 &302\\
   stellar contamination                & 3490  &3490 &0    &0     &0 &0\\
   stellar contamination (stamp)           & 1566  &0    &1566 &0     &0 &0\\
   \hline
\end{tabular}

\end{table}

\section{Classification Method}
\label{sect:classification method}
A CNN was employed for binary classification for stellar contamination case. 
Regarding cloudy contamination situation, three different models, 
ResNet-18, lightGBM, 
and SVM, were individually tested for binary classification.
 The purpose of this four binary classification methods was to distinguish contaminated images from normal images, specifically separating stellar contamination images from normal images and cloudy contamination images from normal images.
\subsection{Stellar contamination}
A CNN is utilized to perform binary classification on the pre-annotated dataset, allowing the network to learn image features and subsequently classify images. CNN, through the combination of convolution operations, padding, pooling, fully connected layers, and activation functions, effectively extracts features from images and conducts advanced pattern recognition. This ability has made CNN widely applicable in computer vision tasks, including image classification, object detection, and image segmentation. In fields such as Astronomy, CNN can also be employed for image processing and pattern recognition tasks, assisting researchers in handling and analyzing celestial images.

\noindent For a large volume of data, CNN can utilize gradient descent to find appropriate parameters, allowing the trained neural network to distinguish specific pattern types effectively.
\subsubsection{CNN Model }
We imported modules such as \texttt{Dense}, \texttt{Flatten}, \texttt{Conv2D}, \texttt{MaxPool2D}, \texttt{Activation} from \texttt{tensorflow.keras.layers} and constructed a CNN structure as illustrated in the Fig.~\ref{Fig1} below.
\begin{figure}[h]
\centering
\includegraphics[height=5cm, angle=0]{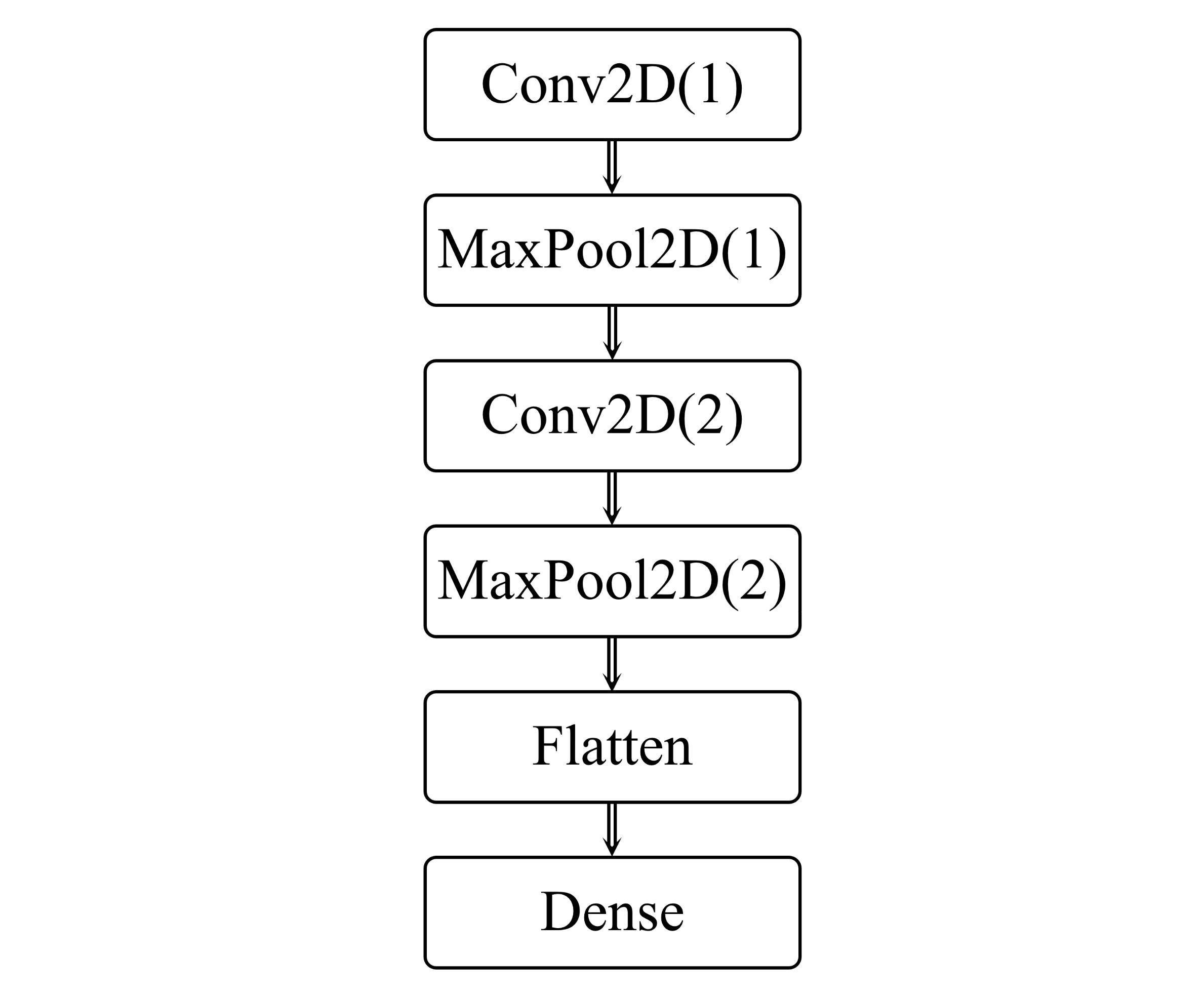}
\caption{CNN structure}
\label{Fig1}
\end{figure}

\noindent The first convolutional layer comprises 5 convolution kernels with a 
size of 4$\times$4 pixels, and the default stride for convolution is 1. The second 
convolutional layer consists of 5 convolution kernels with a size of 3$\times$3 pixels, 
and the default stride for convolution is 1. The ReLU function is employed as the activation 
function. The Dense layer has an output unit count of 1, and the activation function used 
is the sigmoid function. The loss function utilized is 
BinaryCrossentropy (\texttt{tensorflow.keras.losses.BinaryCrossentropy}), and the Adam 
optimizer is applied.

\noindent 90\% of total 8090 images are used as the training set, with the remaining 10\% serving 
as the test set, and a total of 10 epochs were trained. The training results are presented in Section 4.1.
\subsubsection{CNN Model with Stamp Images}
From the example image Fig.~\ref{Fig02}, it can be observed that the class feature of stellar contamination are entirely determined by moving objects and star trails near the moving objects. The remaining parts of the images are considered invalid. Therefore, cropping out the region of interest known as ``stamp image", and training only on these stamp images can enhance classification accuracy.

\noindent Training the CNN with stamp images, the model parameters are the same as those in Section 3.1.1. As expected, there was a significant improvement in model accuracy, as shown in the training results in Section 4.1.

\subsection{Cloudy Contamination}
The ResNet model exhibits low accuracy for the classification of cloudy contamination and normal data. The lightGBM model achieves high accuracy neverless with poor generalization capability. But the SVM model achieves high accuracy with strong generalization ability.
\subsubsection{ResNet Model}

The architecture of deep convolutional neural networks was introduced by \citealt{He+etal+2016} . ResNet primarily addressed the issues of gradient vanishing and exploding in deep neural networks, enabling the training of deeper neural networks. As a result, it achieved significant success in image classification and computer vision tasks. In this study, 
we attempted a ResNet-18 architecture similar to Michael Mommert's (\citealt{Mommert+2020}). We imported modules such as \texttt{Resnet} and \texttt{BasicBlock} from \texttt{torchvision.model.resnet} and used these modules to construct the ResNet-18.

\noindent 80\% of the data is allocated for the training set, and the remaining 20\% is used for the test set. We initiate the optimization process with a learning rate of 0.025, which decreases by a factor of 0.3 every second epoch for a total of 10 epochs. The training and testing results can be found in Section 4.2.
\subsubsection{lightGBM Model}
This section also draws reference from the work of Michael Mommert (\citealt{Mommert+2020}). 
The principles and characteristics of the lightGBM model can be found in the paper. The features of \textit{Time derivatives} employed in the lightGBM model in this study differ from those used by Michael Mommert, since the data used in our study was obtained from the telescope in the moving object tracking mode and has a smaller field of view. The features are as follows:

\noindent \textit{1. Background-related features.} Due to the impact of cloudy contamination on the 
background of FIT images, we utilize three features: median background brightness, mean background brightness, and background brightness standard deviation.

\noindent \textit{2. Time derivatives.} Considering the relative motion of clouds within the 
field of view and the varying sizes of clouds, we opt to calculate the differences between the
 above-mentioned features at the current moment and those from 5 seconds ago, 10 seconds ago, and 15 seconds ago. This results in a total of nine new features.

\noindent \textit{3. Environment features.} The analysis suggests that three features—solar elevation angle, 
lunar elevation angle, and lunar phase—might influence the imaging quality of cloudy layers. 
Actually, it was observed that these three features have minimal impact on the classification results due to the fact that nearly all images are captured at night, and the observation site experiences strong light pollution. In summary, a total of 15 features are considered.

\noindent The decision tree has a maximum depth of 5, with a total of 100 decision trees. Each tree has 20 leaf nodes, and the minimum number of samples on each leaf node is 20. L1 and L2 regularization strengths are set to 30 and 5, respectively. The training results of the model can be found in Section 4.3.
\subsubsection{SVM Model}
SVM is a supervised learning algorithm based on the principle of finding an optimal separating hyperplane in a high-dimensional space to distinguish data points from different categories. The core objective of SVM is to locate a separating hyperplane in such a way that it maximizes the distance between the nearest points of the two different data categories. These closest points are referred to as support vectors.

\noindent In practical applications, data is often not linearly separable. To address non-linear data, SVM employs kernel functions to map the data into a higher-dimensional space, making it linearly separable. The linear kernel, polynomial kernel, and radial basis function are commonly used kernel functions. SVM also involves a regularization parameter that can adjust the margin of the separating hyperplane to prevent overfitting or underfitting the data.
SVM is used for classifying different types of galaxies based on their features and for identifying and categorizing various celestial objects, such as stars, galaxies, quasars, and other cosmic entities.

\noindent GLCM(The Gray-Level Co-occurrence Matrix) is a commonly used tool in image processing and texture analysis to describe the grayscale relationships between pixels in an image. It is possible to obtain textural features, such as \textit{contrast}, \textit{correlation}, \textit{energy}, \textit{entropy} and other texture features, by calculating the GLCM. The calculation formula for the elements of the GLCM is as follows (\citealt{Wang+2019}):
\begin{equation}\label{eq1}
   g(i, j)=\#\left\{f\left(x_1, y_1\right)=i, f\left(x_2, y_2\right)=j \mid\left(x_1, y_1\right),\left(x_2, y_2\right) \in M \times N\right\},
\end{equation}
in which, $x$ and $y$ are coordinates within the image,  $i$ and $j$ are the row and column index of the matrix $g$, $M$ and $N$ are the sum of the rows and columns of image, 
$g$ is the GLCM of the image $f$, $\#$ means the number of elements in the set. 
The distance between $(x_1, y_1)$ and $(x_2, y_2)$ is $d$, and two points' angle 
with the abscissa axis is $q$.

\noindent \cite{Ulaby+1986} found that \textit{1.contrast}, \textit{2.IDM}, \textit{3.energy}, and \textit{4.correlation} are uncorrelated, 
and these four features are easy to compute and provide high classification accuracy. \cite{Baraldi+Panniggiani+1995} conducted a detailed study of six 
texture features and identified \textit{1.contrast} and \textit{5.entropy} as the two most important features. We selected 10 normal 
images and 10 cloudy contamination images then calculated these five features for these 20 images after computing the GLCM with d=1, q=0. 
Additionally, we calculated \textit{6.G}, the grey-value inconsistency of image.
The formulas for these six features are as follows(\citealt{Liu+2003}):

\noindent \textit{1.contrast.} 
\begin{equation}\label{eq1}
   \mathrm{CON}=\sum_{i=1}^{M_g} \sum_{j=1}^{N_g}(i-j)^{2}g(i, j)
\end{equation}
\noindent $M_g$ and $N_g$ represent the total number of rows and columns in matrix $g$.

\noindent \textit{2.Inverse Difference Moment.} 
\begin{equation}\label{eq2}
   \mathrm{IDM}=\sum_{i=1}^{M_g} \sum_{j=1}^{N_g}\frac{g(i, j)}{1+(i-j)^2}
\end{equation} 
\noindent \textit{3.energy.} 
\begin{equation}\label{eq3}
   \mathrm{ENE}=\sum_{i=1}^{M_g} \sum_{j=1}^{N_g}g(i,j)^2
\end{equation}
\noindent \textit{4.correlation.} 
\begin{equation}\label{eq4}
   \mathrm{COR}=\sum_{i=1}^{M_g} \sum_{j=1}^{N_g}\frac{(i-\mu)(j-\mu)g(i,j)^2}{\sigma^2}
\end{equation}
\noindent \textit{5.entropy.} 
\begin{equation}\label{eq5}
   \mathrm{ENT}=-\sum_{i=1}^{M_g} \sum_{j=1}^{N_g}g(i,j)log(g(i, j))
\end{equation}
\noindent \textit{6.G.} The grey-value inconsistency of image is an important index to represent the image's background grey value, the image's grey-value inconsistency is defined as (\citealt{Wang+2019}):
\begin{equation}\label{eq6}
   \mathrm{G}=10lg(\frac{P_s}{P_n}),
\end{equation}
in which, $P_s$ and $P_n$ are respectively the maximum and minimum standard deviation of local images. 
This study selects the template of $9\times9$ pixels to spread all over the image.

\noindent After normalizing the computed feature values to a range between 0 and 1, it was observed that these 6 features can effectively distinguish between the two categories, as shown in the Fig.~\ref{Fig2} below. Therefore, this study will utilize these six features for training the SVM model. 
The SVM model training and testing results are displayed in Section 4.4.
\begin{figure}[h]
   \centering
   \includegraphics[height=8cm]{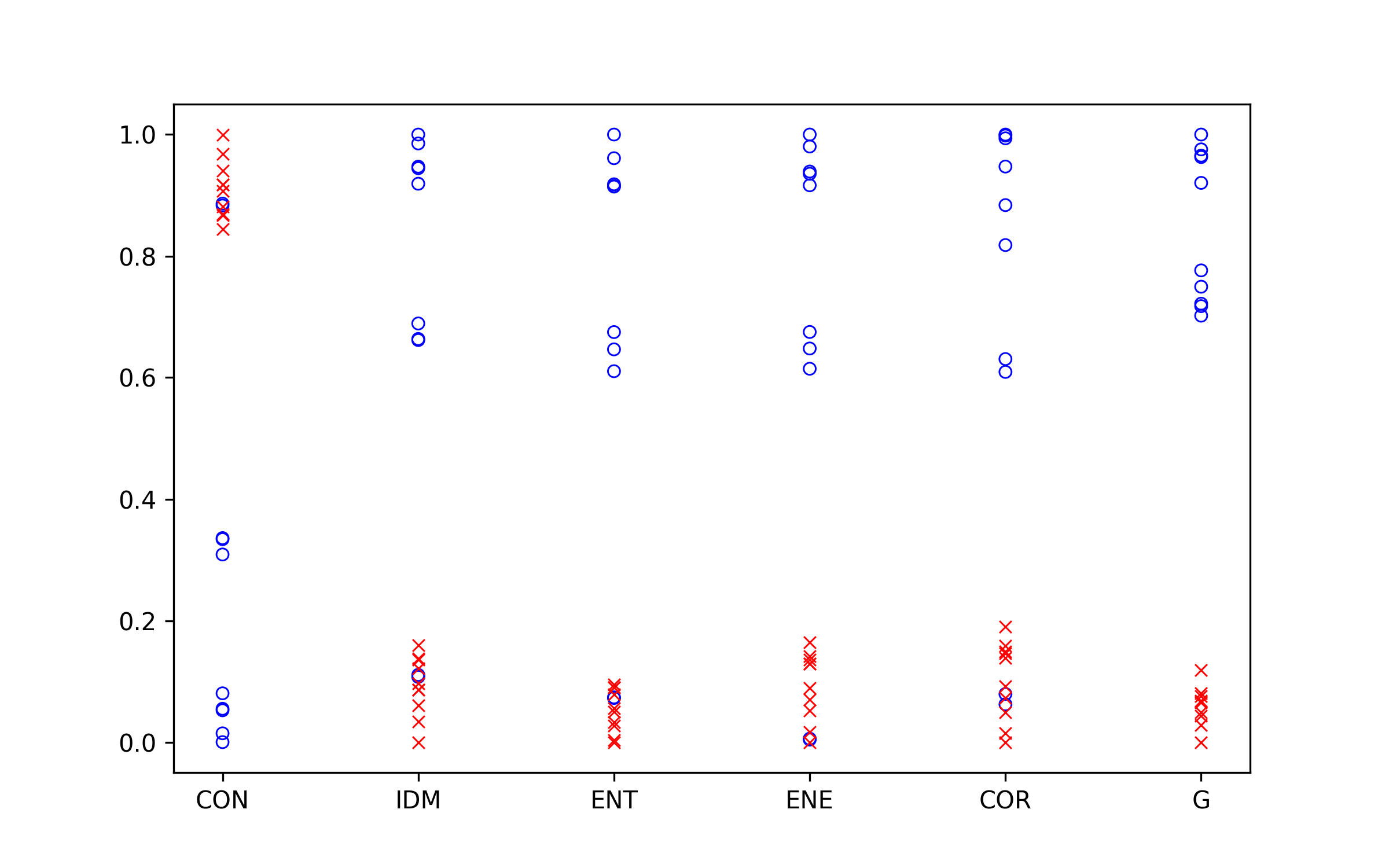}
   \caption{The distribution of the 6 features for the 20 images. Blue: normal images, while red: cloud-contaminated images.}
   \label{Fig2}
\end{figure}

\section{result}
\label{sect:result} 
In this section, the results for the four mentioned models are presented. 
Tab.~\ref{Tab2} below shows the accuracy of these models on the test set and the training set.

\begin{table}[h]
   \centering
   \caption[]{Accuracies on the training set and test set of all four models.}\label{Tab2}
   
   \begin{tabular}{clcc}
   \hline\noalign{\smallskip}
   Contamination & Model & Training set & Test set \\
   \noalign{\smallskip}\hline
   \multirow{2}{*}{stellar} & CNN & 0.9389 & 0.9221 \\
                         & CNN with stamp images & 1.0000 & 1.0000 \\
   \hline
   \multirow{3}{*}{cloudy} & ResNet & 0.8000 & 0.8000 \\
   & lightGBM & 1.0000 & 0.9995 \\
   & SVM & 0.9753 & 0.9712 \\
   \hline
   \noalign{\smallskip} 
\end{tabular}
\tablecomments{0.86\textwidth}{The accuracies of LihgtGBM model shown above are very high, 
but this model has weak generalization ability.}
\end{table}

\subsection{Identify Stellar Contamination}
\subsubsection{Training and Testing CNN Model}
\begin{figure}[h]
   \begin{minipage}[t]{0.495\linewidth}
   \centering
    \includegraphics[width=60mm]{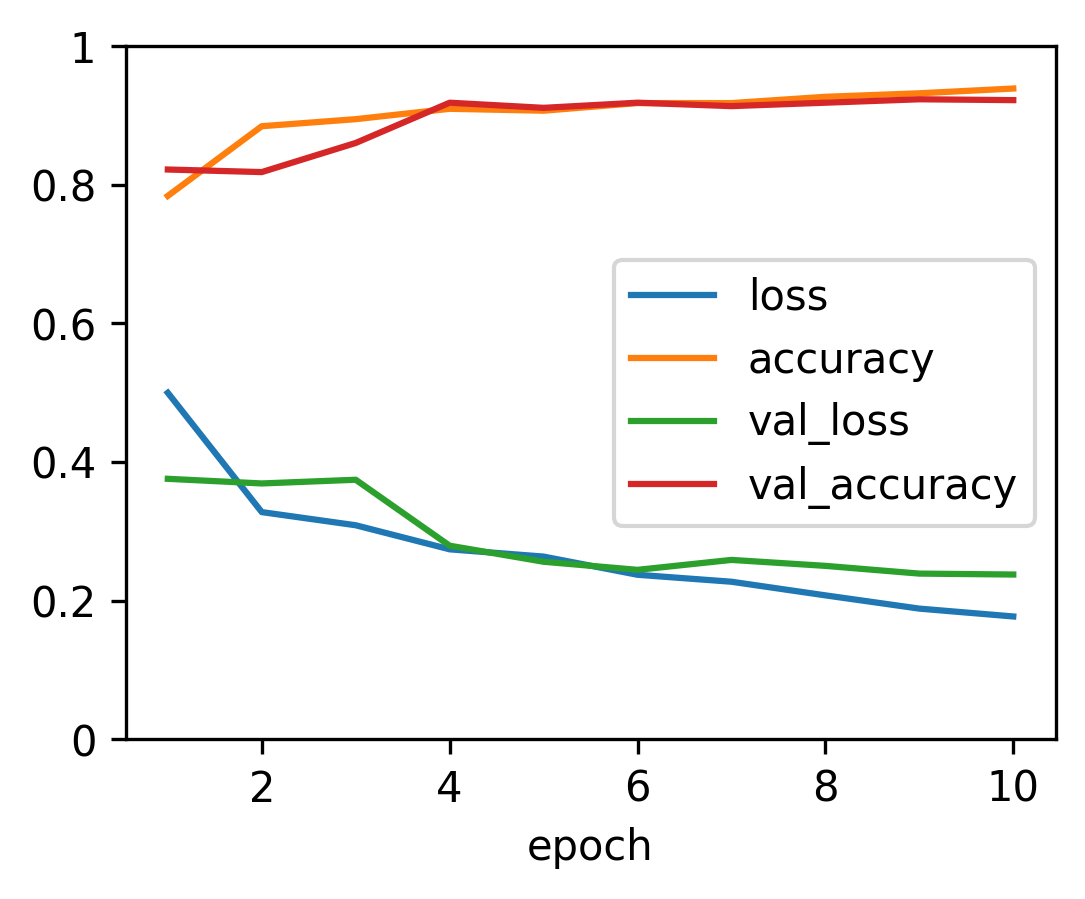}
      \caption{\label{Fig3}{\small Growth curves of CNN model} }
   \end{minipage}%
   \begin{minipage}[t]{0.495\textwidth}
   \centering
    \includegraphics[width=60mm]{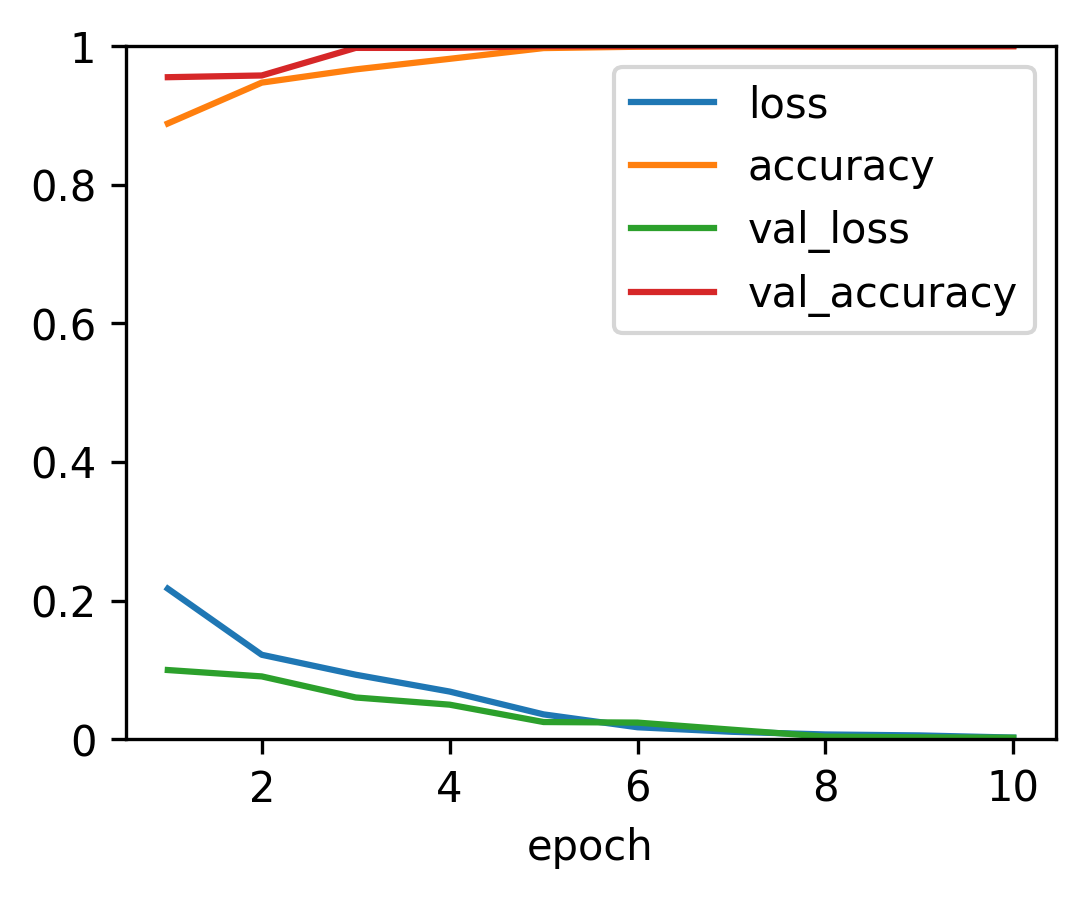}
      \caption{\label{Fig4}{\small Growth curves of CNN model with stamp images}}
   \end{minipage}%
 \end{figure}
 From the growth curve in the above Fig.~\ref{Fig3} and Fig.~\ref{Fig4}, 
 it can be observed that the CNN model trained with stamp images converges 
 faster and exhibits better performance on the test set. We import module \texttt{f1\_score} from \texttt{sklearn.metrics} and use it to calculate F1 scores. The F1 scores for
  this two different cases are 0.90 and 1.00, respectively. Their confusion matrices on the test set are illustrated in the Fig.~\ref{Fig5} and Fig.~\ref{Fig6} below.
  Confusion matrix is a visual tool designed for supervised learning. 
  In confusion matrix, each column represents the predicted resluts, while each row corresponds to the truth. ``Positive" represents contaminated images, while ``Negative" represents normal images.
  In Fig.~\ref{Fig5}, CNN model incorrectly classified 51 contaminated images as normal images and 12 normal images as contaminated images. In Fig.~\ref{Fig6}, CNN model trained with stamp images made all correct predictions.
\begin{figure}[h]
\begin{minipage}[t]{0.495\linewidth}
\centering
   \includegraphics[width=65mm]{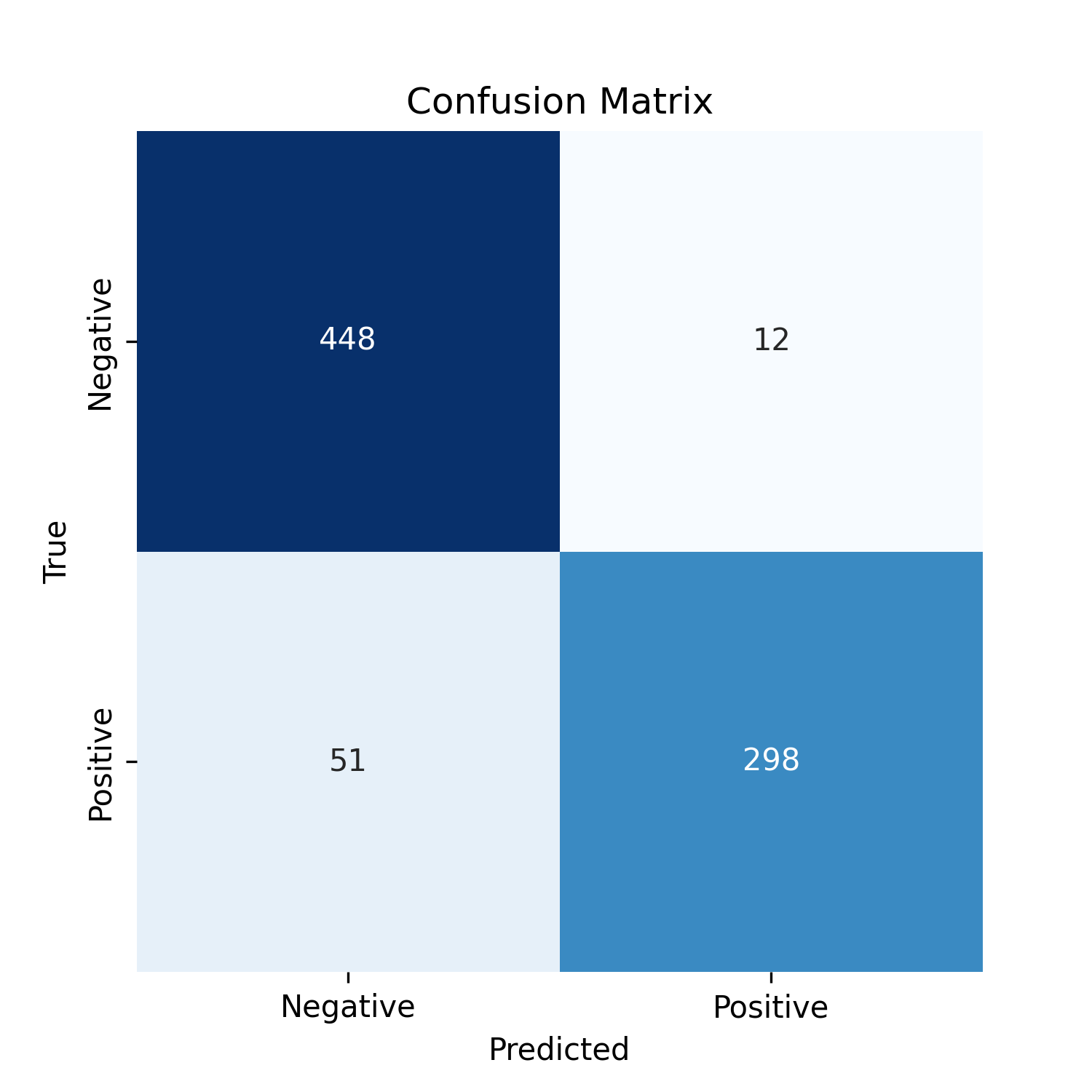}
   \caption{\label{Fig5}{\small Confusion matrix of CNN model} }
\end{minipage}%
\begin{minipage}[t]{0.495\textwidth}
\centering
   \includegraphics[width=65mm]{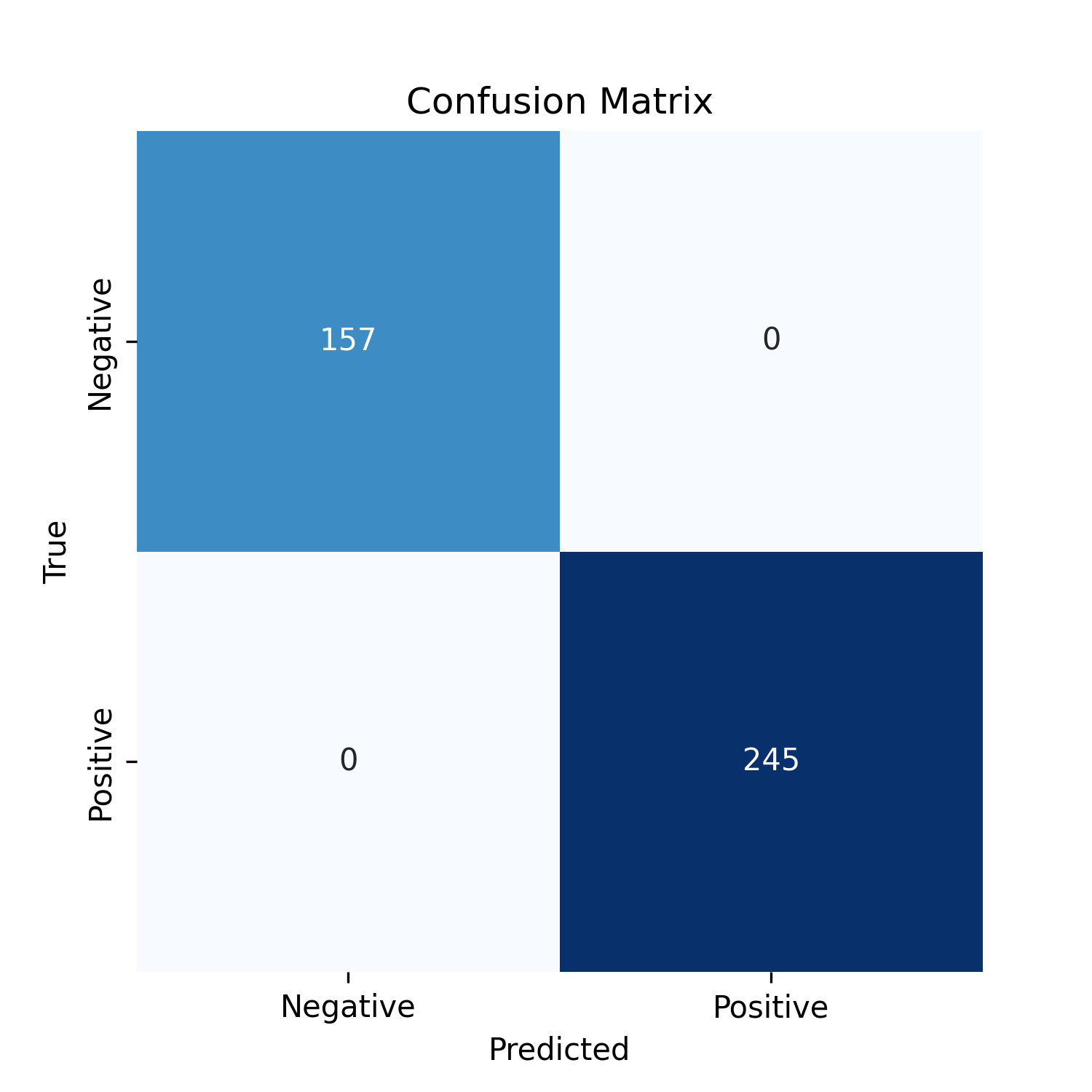}
   \caption{\label{Fig6}{\small Confusion matrix of CNN model with stamp images}}
\end{minipage}%
\end{figure}

\subsection{Identify Cloudy Contamination}
\subsubsection{Training and Testing ResNet-18}
\begin{figure}[h]
   \centering
   \includegraphics[width=60mm]{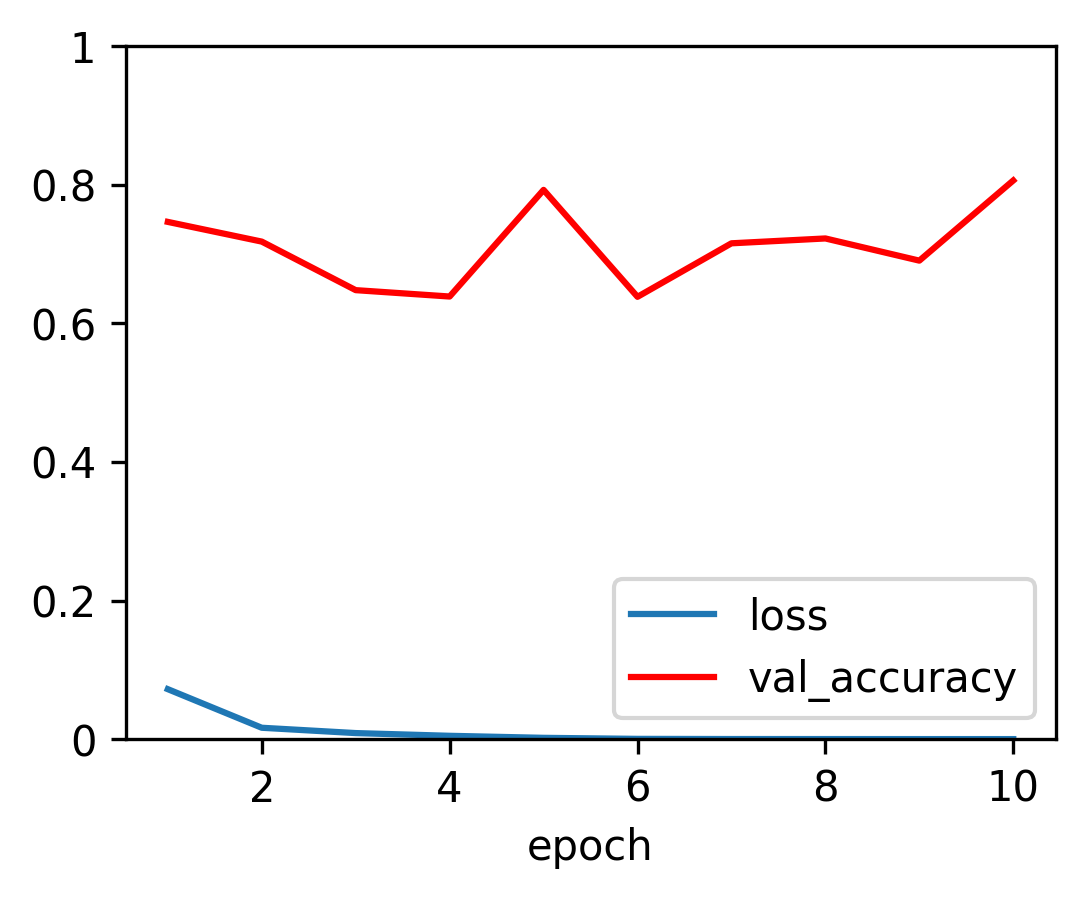}
   \caption{Growth curve of ResNet-18 model}
   \label{Fig6.1}
\end{figure}

The model's accuracy on the test set is below 0.8 with epoch being set to 10, as shown in Fig.~\ref{Fig6.1}, which is lower than Mommert's accuracy of 0.85. Our analysis suggests that this could be due to the smaller field of view in the images and the relatively faster motion of clouds within the field of view, resulting in less distinct cloudy features.
\subsubsection{Training and Testing lightGBM Model}
\begin{figure}[h]
\begin{minipage}[t]{0.495\linewidth}
\centering
   \includegraphics[width=65mm]{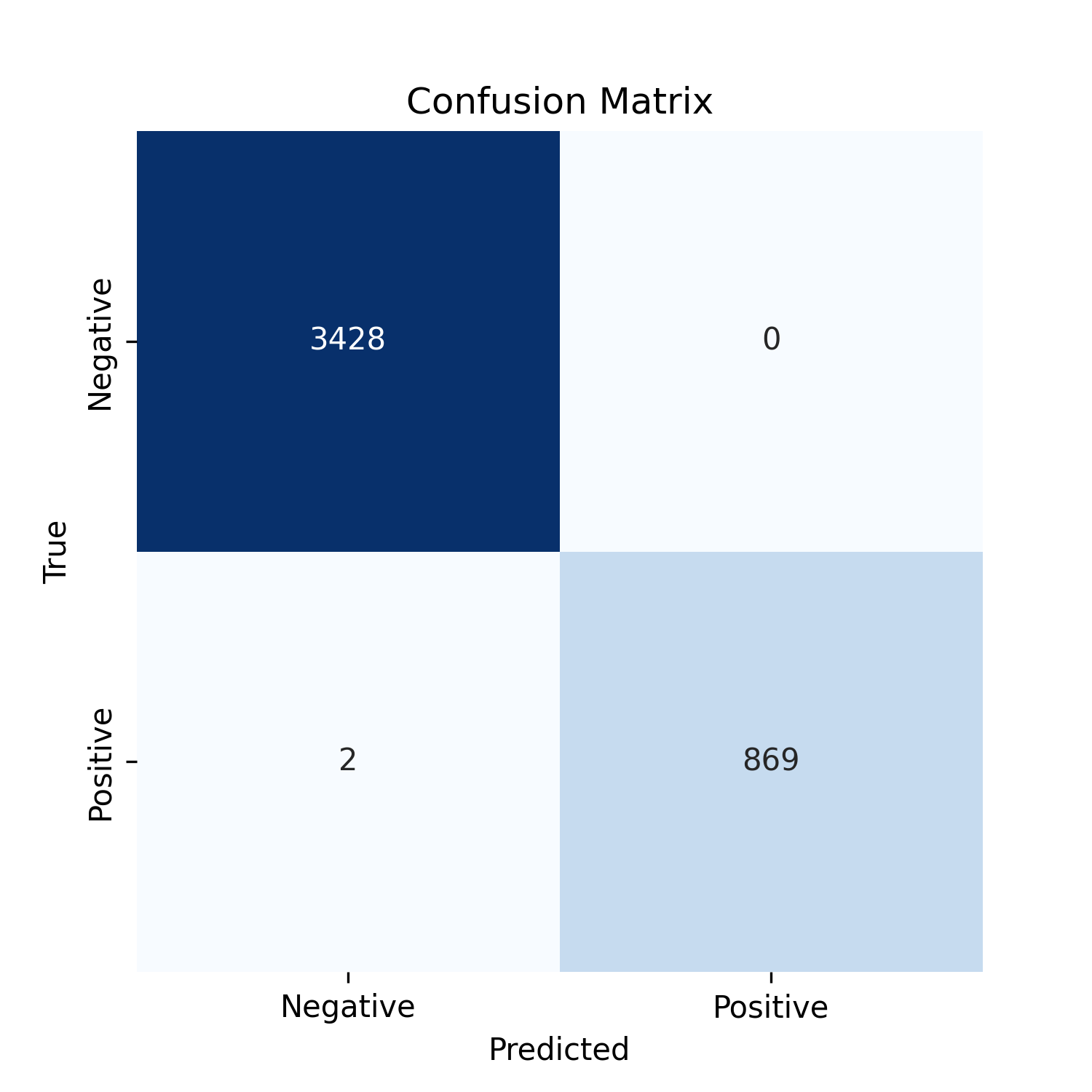}
   \caption{\label{Fig7}{\small Confusion matrix } }
\end{minipage}%
\begin{minipage}[t]{0.495\textwidth}
\centering
   \includegraphics[width=65mm]{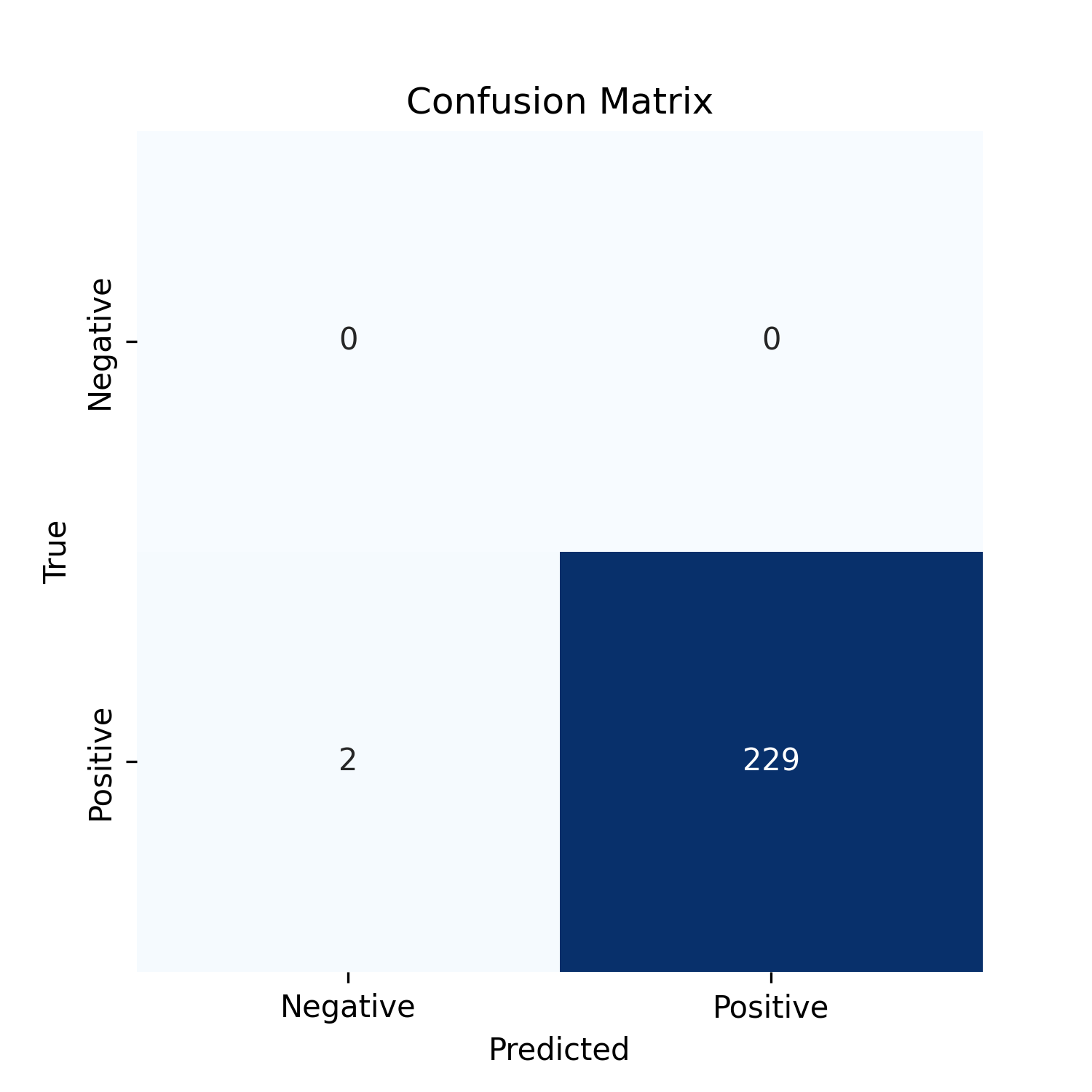}
   \caption{\label{Fig8}{\small Confusion matrix on data of October 12}}
\end{minipage}%
\end{figure}

The lightGBM model, when trained with data from October 4, 2022, achieved an impressive classification 
accuracy of 99.95\%, as shown in the confusion matrix in Fig.~\ref{Fig7}.
However, it displayed poor generalization with nearly zero accuracy when this model was used to classify data from October 12. 
But when this 2 days' data were combined for model training, the consequent model performed well when applied to classify images from October 12 again, as indicated in the confusion matrix in Fig.~\ref{Fig8}. It mistakenly recognized only 2 contaminated images as normal.

\noindent The analysis suggests that learning cloudy imaging features from data on a single day or a few days may not be able to capture the general characteristics of cloudy layers. Therefore, long-term data accumulation is needed to improve the generalization ability of the lightGBM model.

\subsubsection{Training and Testing SVM Model}
\begin{figure}[h]
\begin{minipage}[t]{0.495\linewidth}
\centering
   \includegraphics[width=65mm]{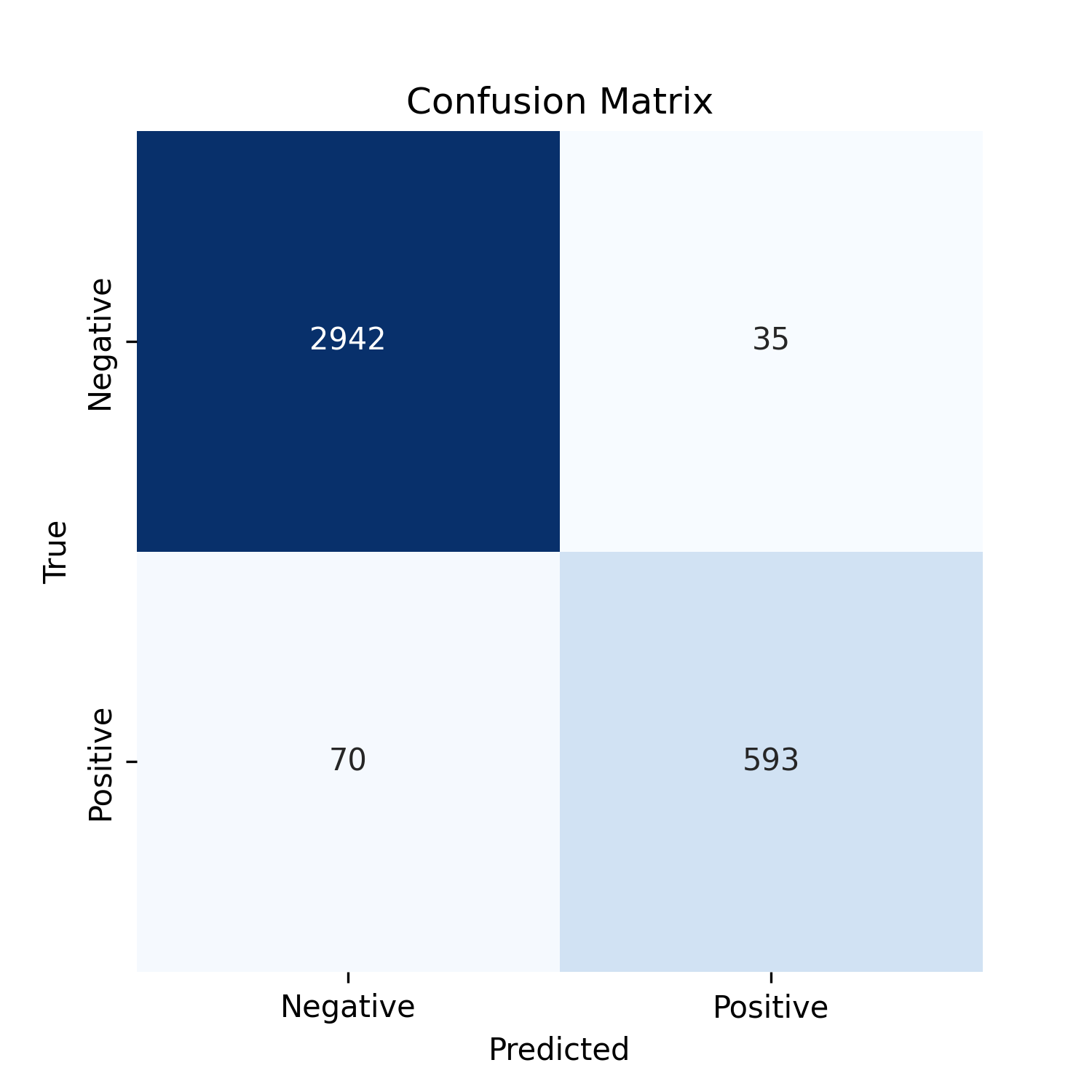}
   \caption{\label{Fig9}{\small Confusion matrix } }
\end{minipage}%
\begin{minipage}[t]{0.495\textwidth}
\centering
   \includegraphics[width=65mm]{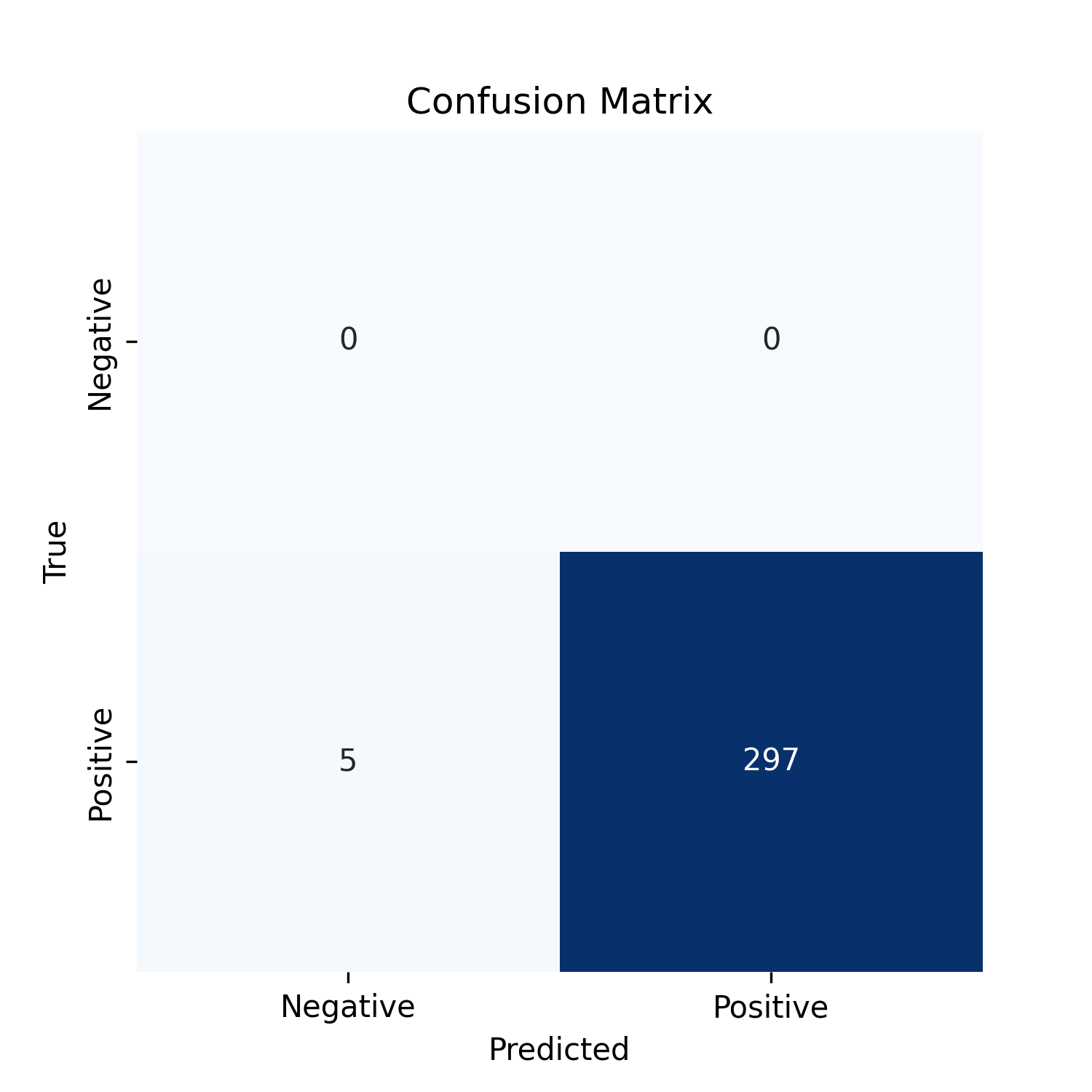}
   \caption{\label{Fig10}{\small Confusion matrix on data of October 12}}
\end{minipage}%
\end{figure}

The SVM model, trained with 15,000 samples of normal data and 3,199 samples(October 4) of 
cloudy contamination data, achieved an accuracy of 97.12\% on the test set. This model's 
accuracy for classifying data from October 12 remains high at 98.34\%, indicating that the 
SVM classification model exhibits good generalization ability. The confusion matrices for 
the SVM model on the test set and the data from October 12 can be found 
in Fig.~\ref{Fig9} and Fig.~\ref{Fig10}, respectively. The recall rate are 89.44\% and 98.34\%.

\section{Conclusions}
\label{sect:conclusion}
\noindent A dataset for machine learning training has been obtained through manual annotation of
 a portion of the YNAO's 2022 observational data and additional data from 2023. The dataset with total 40,115 images is of high quality, well-classified, and easy to use.

\noindent From the results of the two CNN models, it's evident that the CNN model trained with stamp
 images achieves higher accuracy around 100\%, meeting the practical requirements to identify stellar contamination from normal images. 

\noindent Among the three models for identifying cloudy contamination, ResNet-18 has lower 
accuracy on both training and testing dataset, but the other two models perform well. The lightGBM model exhibits poor generalization 
and requires long-term data accumulation to overcome, while
the SVM model has stronger generalization, maintaining high classification accuracy with 97.12\% for new samples.

\begin{acknowledgements}
This work was funded by the funds of National Natural Science Foundation of China (No. 12373086, No. 12303082), CAS ``Light of West China" Program, Yunnan Revitalization Talent Support Program in Yunnan Province, and National Key R\&D Program of China, Gravitational Wave Detection Project No.2022YFC2203800.
Thanks to He Zhao from Purple Mountain Observatory for the assistance. 
Thanks to Michael Mommert from Lowell Observatory for the open-source code.
\end{acknowledgements}

\label{lastpage}

\end{document}